\documentclass[twocolumn,prX,longbibliography]{revtex4}
\usepackage{braket}
\usepackage{amsmath}
\usepackage{graphicx}
\usepackage{amsfonts}
\usepackage{float}
\usepackage{mwe}
\usepackage{subcaption}
\begin{document}

\title{Quantum Projective Simulation with Hamiltonian Evolution: A study in reinforcement learning}
\author{Amara \surname{Katabarwa}}
\email{akataba@uga.edu}
\author{ Nima \surname{Karimatari}}
\email{nima.kt@uga.edu}
\affiliation{University of Georgia}
\date{\today}

\begin{abstract}
Projective Simulation was introduced as a novel approach to Artificial Intelligence. It involves a deliberation procedure that consists of  a random walk on a graph of clips and allows for the learning agent to project itself into the future before committing to an action. Here we study and analyze a quantum mechanical version in which the random walk is performed by two kinds of Hamiltonians. The first kind is implemented by naively embedding the classical model in a quantum model by turning the clips into qubits. The other allows for storing clips in superpositions of qubits allowing for a potentially purely quantum mechanical learning procedure in which the perception of the environment is purely quantum mechanical but the action is classical. We lastly introduce the concept of interacting projective agents for both the classical and quantum mechanical case.
\end{abstract}
\maketitle

\section{Introduction}

In this work we study model of artificial intelligence introduced in \cite{J.Briegel2012} that is a version of reinforcement learning. According to \cite{Sutton1998}, reinforcement learning is, ``learning to   what  to  do-how  to  map  situations  to actions-so  as  to  maximize  a  numerical  reward  signal''. A key feature of reinforcement learning is that the learning agent is not ``told'' what actions are the right ones since the goal is for the agent to learn what to do in novel external environments where no model is given. Although projective simulation can be seen as a version of reinforcement learning there are important ways in which it does not quite fit this paradigm. As a consequence of this, we briefly describe reinforcement learning in order that the important difference maybe more easily seen. 

\subsubsection*{Elements of reinforcement learning}

\textit{A policy} is a way of determining the agent's action predicated on it's perception of the states of the environment. More formally it will be a mapping from the set of states and actions of the agents to a conditional probability \( \pi(a|s) \) where \(a\) is the action taken and \(s\) is the state of the agent.\newline
\indent \textit{A reward function} is a mapping from state-action pair to a real number \(r(a|s)\). It represents the desirability of an action. The goal of the learning agent is to maximize the long term reward. \newline
\indent Simply considering immediate rewards may not in the long run lead to maximizing the reward.For this purpose reinforcement also has a \textit{value function}. This assigns a real number to the total reward an agent expects to get in the long run given likely states to follow and the rewards for those states. \newline
\indent The last feature of reinforcement learning is the environment. The environment is largely anything the agent can't arbitrarily change.\newline
\indent We now formulate more precisely what we mean by maximizing the expected reward before introducing the projective agent.From now on we assume the learning process is a Markovian Decision Process(MDP) so that there are a finite number of states and actions available for the agent\newline
For each time step \(t\) we have an reward represented by \(R_t\). The total reward then for time steps \(t_0\) to \(t\) is then
\begin{equation}
 H'_t = R_o + R_1 + \dots R_t + \dots R_T,
\end{equation}
where \(T\) is the final time being considered. \(T\) can be made to be infinity or some finite value. We want the agent to have the flexibility of being able to consider just immediate rewards or rank more subtlety future rewards. We therefore introduce a parameter \( \gamma \) and re-write the total reward as
\begin{equation}
  H_t = \sum_{k=0}^{T-t-1} \gamma^k R_{t+k+1}.
\end{equation} 
Note that when \( \gamma =0 \) only the reward for immediate action is accounted for while \(\gamma = 1\) accounts for all the rewards for all time steps being considered. In general \( 0 \geq \gamma \geq 1 \). We then define a value function \(v_{\pi}\) which is the expectation of the total reward. In general.

\begin{equation}
 v_{\pi}(s) = \mathbb{E}_{\pi}[H_t|s]=  \mathbb{E}_{\pi}[\sum_{k=0}^{k= \infty} \gamma^k R_{t+k+1}|s].
\end{equation}

In the equation above \(\pi\) is the policy being implemented.The goal of the learning procedure is to find the optimal policy \( \pi_{*} \) that will maximize the value function \(v_{\pi}(s)\).

\subsubsection*{Projective Simulation}
Projective Simulation has the same basic features of reinforcement learning in the sense that we have an agent with possible states and actions with no access to the model of the environment. It gets penalized or rewarded for specific actions in other words we have a policy which is stochastic. It departs from the reinforcement paradigm in the sense that we do not have a value function that takes into account future rewards.In a sense the lack of a value function is replaced by a structure of called the \textit{episodic and compositional memory} (ECM). The components of a projective agent are the following:
\begin{itemize}
 \item \textit{ECM} : This is a directed and weighted graph of clips
 \item \textit{Clip}:These represent episodic experiences. In general a clip is an n-tuple \(c =(c^{1}, c^{2}, \dots c^{n})\). Each \(c^{i}\) is either a perception or an action.
 Note: an episode is a length of finite period of time which usually ends in an action
 \item \textit{Percepts}: This is an m tuple \(s =(s^{1}, s^{2}, \dots s^{m}) \in \mathcal{S} \equiv S^1 \times S^2 \times  \dots S^m \). The tuple represents the kind of data that the agent can perceive. For example, \(S^1\) could represent the set of colors the agent can represent while \(S^2\) can be the set of shapes the agent can perceive. The \(m\) tuple is then the collection of perception or inputs
 \item \textit{Actions}: This is an \(l\) tuple \(a = (a^{1}, a^{2},\dots a^{l}) \in \mathcal{A} \equiv A^1 \times A^2 \times \dots A^{l}\). This represents the sets of actions available to the agent
 \item \textit{Emotional Tags}: These are markers that can be associated to edges of clips to indicate for example whether the transition between the considered clips was rewarded. This is the projective agent's way of moving beyond simply rewarding only immediate actions. Formally an emotional tag is \( e(s,a)= (e^1,e^2, \dots e^k) \in \mathcal{E} \equiv E^1 \times E^2 \times \dots E^k \). Using the mechanism of \(reflection\) the agent can exploit the emotion tangs before an action is taken as a response to being in a certain state.
 \item \textit{Reward:} \( h^t(c_i, c_j)\). Each edge between clip has a weight \( h^t(c_i, c_j)\) that strengthens transitions between clips \(c_i, c_j\) as the value of \( h^t(c_i, c_j)\) gets bigger. This will be referred to as the \textit{weight} matrix. It is initially initialized to an arbitrary value
 
\end{itemize}

The learning process for the projective agent is:
\begin{enumerate}
 \item Each time step begins with a percept being excited by the environment i.e exciting a memory clip \(c_i\).
 \item The excitation hops to clip \(c_j\) with probability
  \(p^t(c_i\j|c_i) = \frac{h^t(c_i, c_j)}{\sum_k h^t(c_i, c_k)} \) where \(c_k\) is any clip connected to \(c_i\) on the way to \(c_j\).
  \item At some time step the projective agent performs the action which is either reward or not rewarded. The weight matrix is then updated by the following rule
  \begin{equation} 
  \begin{split}
      h^{t+1}(s, a) =  h^{t}(s,a) - \gamma[ h^{t}(s,a)-1] +\\ \lambda \delta(s,s^t)\delta(a,a^t).
  \end{split}
  \end{equation}
\end{enumerate}
The notation for the clips has been changed to emphasize the transition from a state of the learning agent to an action performed. Here \( \alpha \) is a dissipative factor and \(\lambda \) is the reward given when the action taken is the right one.
\newline
\indent As can be seen the projective agent in the simplest case is far more primitive than the traditional reinforcement learning agent. More bells and whistles can be added to \( ECM\) to make the learning procedure of the agent more attentive and responsive to the environment, like dynamically adding more clips and making new associations (adding new edges) between clips \cite{J.Briegel2012}
The classical projective agent so far has been tested in a different number of situations and been shown to eventually learn what needs to be done \cite{M.Tiersch2015,Friis2015,Melnikov2015} 

\section{Quantum Projective Agent:\newline Two Models}
 \subsection{First Model}
In \cite{J.Briegel2012}, the first quantum projective agent was defined by a Hamiltonian. The \(ECM\) is accounted for by a tensor product in the Hilbert space of clips.  So that 
\begin{align}
 c_i \longmapsto \ket{c_i} ,
\end{align}
with the classical random walk being replaced by a quantum walk and hopping probabilities
\begin{align} 
	p(c_i|c_j)\longmapsto
	 \vert\braket{c_i|c_j} \vert^2,
\end{align}  for single hop between neighboring clips and for hops among a number of different clips we have

\begin{align}
 p(c_j|c_i) \longmapsto \sum_{k} \vert\braket{c_j|c_k}\braket{c_k|c_i}\vert^2 ,
\end{align}

We now have a graph \( \mathcal{G}=(V,E) \) with vertices
\(j\in V \) that label a clip \(c_j \in C\) and  have an edge \( \{j,k\} \in E \) between clips \(c_j \) and \(c_k\).
A quantum walk is then governed by the following Hamiltonian

\begin{equation}
 H = \sum_{\{j,k\}\in E} \lambda_{jk}\left( \hat{c}_k^{\dagger}\hat{c}_j + \hat{c}_k \hat{c}_j^{\dagger} \right) + \sum_{j\in V} \epsilon_j \hat{c}_j^{\dagger}\hat{c}_j,
\end{equation}

where the operators \(\hat{c}_j, \hat{c}_j^{\dagger}\) have the following properties
\begin{align}
 \ket{c_j} = \hat{c}_j^{\dagger}\ket{vac}, \\
 \ket{vac} = \hat{c}_j \ket{c_j},
\end{align}

Where necessary the operator \(\hat{s_i}^{\dagger} \) and its hermitian conjugate will refer to exciting or de-exciting a \(i^{th}\) percept clip while \(\hat{a_i}^{\dagger} \) will refer to exciting or de-exciting an \(i^{th}\) actuator clip.

The Hamiltonian above induces coherent transitions but we can also have incoherent transition with the following dissipative operator.

\begin{equation}
 L\rho = \sum_{\{j,k\}\in E} \kappa_{jk}\left(\hat{c}_k^{\dagger}\hat{c}_j \rho \hat{c}_k \hat{c}_j^{\dagger}- \frac{1}{2}[\hat{c}_k^{\dagger}\hat{c}_j\hat{c}_k\hat{c}_j^{\dagger},\rho]_{+} \right) ,
\end{equation}
where \([-,- ]_{+}\) is the anti-commutator and \( \rho \) is the density matrix of the \(ECM\). \newline
 Recall in general a clip \(c\) is an \(l\)-tuple i.e \(c= (c_1, c_2, \dots, c_l)\) with \(c_{i}=s_i \) or \(c_i = a_i\). Therefore the quantum state corresponding to this clip is given by
 
 \begin{equation}
  \ket{c} = \hat{c}^{\dagger}_1 \otimes \hat{c}^{\dagger}_2 \otimes \dots \otimes \hat{c}^{\dagger}_l \ket{vac}.
 \end{equation}.
 
 The observables that one calculates are the probability of finding an excitation in a quantum state representing a certain action i.e
 
 \begin{equation}
 tr(\hat{a_i}^{\dagger}\hat{a_i}\rho),
 \end{equation}
 
 where \(\hat{a}^{\dagger} \) inserts an excitation representing an action taken by the quantum projective agent.
 
 The learning process will then constitute applying the classical update rule to the coupling in the Hamiltonian i.e
   \begin{equation}
   \begin{split}
       \lambda^{t+1}_{jk} =  \lambda^{t}_{jk} - \alpha[ \lambda^{t}_{jk}-1] +\\ \gamma \delta(j,k^t)\delta(j,k^t).
   \end{split} 
   \end{equation}
 
 In later work \cite{Paparo2014} another version of a quantum projective agent was introduced that used Szegedy walk operator \cite{Szegedy2004} which fits into a Grover-like procedure to do the quantum walk. It was shown in that work that there is a square root speed up for a version of that quantum projective simulation called  \textit{Reflecting Projective Simulation} . \newline
 In the original work a learning procedure was never explicitly laid out and it was never demonstrated that indeed the quantum projective agent can indeed learn. This is what we demonstrate in this work. \newline
 \indent Also one might wonder why bother with Hamiltonian evolution especially since there is already a version of the projective agent with a quantum speed up. A possible reply is the following: one of the problems faced today is sometimes not necessarily speed up but data storage. As a consequence the data to be search is usually stored across different networks which might turn out be in different physical locations. When this is the case it is hard to see how Grover type methods can be applied since it requires a superposition of all the elements in one's database \cite{Viamontes2005}. Therefore the issue of storage may be one that is more amenable to Hamiltonian evolutions. One could imagine using heralding techniques like those envisioned for ion trap quantum computing \cite{Brown2016}. \newline
 \indent In order to fully realize this desire we would like a Hamiltonian that more efficiently uses the Hilbert space. For the case where the clips \(c\) are length one i.e \(c = s\) or \(c=a \) and in which excitations of percepts directly leads to actions, we introduce a new Hamiltonian where the percepts are stored in superpositions of quantum states.  \newline
 
 \subsection{Second Model}
 We need a better use of the Hilbert space than is used in first model. For the first model, an excitation of a percept clip is done in the following manner
 \begin{equation}
   \hat{s}^{\dagger}\ket{0}= \ket{1}.
 \end{equation}
 A network of two clips \( (p_0, p_1)\) would then be 4 dimensional Hilbert space so that on this network of two percepts the state  \(\ket{01} \) would mean the second percept being excited and the first not. Note though that the presence of the excitation is what we really care about and we need not encode the fact that other clips are not excited. This suggests a different encoding is possible. The degrees of freedom of the qubits will be mapped to presence of an excitation corresponding to a percept. Thus, if \(p_0 \) is excited, this is represented by \( \ket{0} \) but if \( p_1 \) is excited the is represented by \( \ket{1} \) \newline
 
 We have thus encoded two classical percepts into one qubit. The quantum state \( \ket{\Psi}= \alpha\ket{0} + \beta \ket{1} \) represents the possibility of exciting either state. The probability of exciting the \( p_0 \) is represented by \( \vert \alpha \vert ^2 \) while the probability of exciting \( p_1\) is given by \( \vert \beta \vert^2 \) . 
 
 As an example consider a network of 3 qubits. We desire to encode a classical agent with 2 percepts and two actuators. The percepts labeled by \( p_0, p_1 \) and the actions labeled by \(a_0, a_1 \). Take two possible events, the first being excitation of \(p_0\) then \(a_1\) denoted by \( (p_0,a_1) \) while the second being the excitation of \(p_1\) then \(a_0\) denoted by  \( (p_1, a_0)\). We then have the following mapping between the classical states and quantum states
 
 \begin{align}
  (p_0,a_1) \longmapsto \ket{001} \\
  (p_1, a_0) \longmapsto \ket{110}
 \end{align}

\textbf{Note:} We have two qubits to represent the presence of two possible classical states of the agent's action and the presence of the excitation in one or the other represents one action or the other. This means that we do not have superpositions of qubits representing actions. We do this merely to keep simplicity of the model. \newline
 
The proposed new Hamiltonian is the following:
 
 \begin{equation}
  H = \sum_{p \in\mathcal{P},a \in \mathcal{A} }\sum_{j=0}^{j=1}  \ket{j}\bra{j}_{p_j} \lambda_{p_ja} \left(c_a^{\dagger} + c_a\right)  + \sum_{i} \epsilon_i c_i^{\dagger}c_i
 \end{equation}
 
 Where \(\mathcal{P} \) is the set of quantum density matrices that correspond to the quantum percepts(quantum states that encode the classical percepts) i.e \( p\) is the quantum percept and \( p_i\) is the classical percept being encoded. \(\mathcal{A} \) is the set of quantum percepts that encode the classical actuators.

  A generalization to qudits can be easily imagined so that the Hamiltonian becomes 
 \begin{equation}
  H = \sum_{p \in\mathcal{P},a \in \mathcal{A} }\sum_{j=0}^{j=d-1}  \ket{j}\bra{j}_{p_j} \lambda_{p_ja} \left(c_a^{\dagger} + c_a\right)  + \sum_{i} \epsilon_i c_i^{\dagger}c_i.
 \end{equation}

We then have the observables of the following type to  measure 
 \begin{align}
  P_{p_ia_k} = I_p \otimes \hdots I_p \otimes \ket{0}\bra{0}_{p_i}\otimes \hdots \otimes \ket{1}\bra{1}_{a_k} \otimes \hdots \ket{0}\bra{0}, \\
    P_{p_ja_l} = I_p \otimes \hdots I_p \otimes \ket{1}\bra{1}_{p_j}\otimes \hdots \otimes \ket{0}\bra{0} \otimes \hdots \ket{1}\bra{1}_{a_l}, 
  \end{align}
  where \(p\) merely means percept qubit while \(p_i\) labels the \(p_i\) classical percept that is encoded in that qubit and \(a\) means actuator. The above are merely two examples. 
  \section{Toy Model and Results}
  
  To demonstrate the learning process for the two quantum models we use a 4 clip network with percept clips \(p_0, p_1 \) and actuator clips \(a_0, a_1\).The goal is simple; the agent must learn that if \(p_0\) is excited then the right action is to excite \(a_0\) and similarly if \(p_1\) is excited then the right clip to excite is \(a_1\). Initially each percept clip is has an edge to each actuator clip and there are no edges between different percepts or edges between different actuators. \newline 
  \indent This is sometimes called the invasion game since one can imagine an attacker and a defender on different sides of a wall with two doors. Before the attacker moves it shows a sign indicating whether it will move left or right. The defender is supposed to learn the meaning of the signs so that when it sees a  ` move right' sign it moves right and defends the attack. Consequently, we call the efficiency of the learner the `learning efficiency'. A learning efficiency of 1 means perfect defense while that of 0 means the agent has learned precisely the wrong thing.
\begin{center}
   \begin{figure}[H]
     \includegraphics[height=65mm, width=65mm]{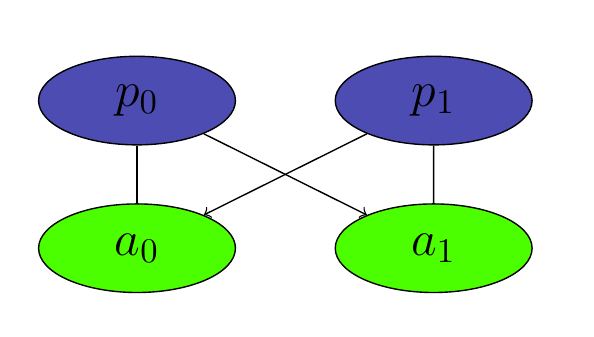} 
     \caption{\(ECM\) for agent}
   \end{figure}
\end{center}   

Since the probabilities for the given Hamiltonians will oscillate, we pick the first instant in time at which the probability of the observable is at its maximum \cite{Johansson2013}.

\subsection*{Interacting Agents}

As we have already mentioned one of the main purposes of reinforcement learning is to create an agent that can interact with the environment (anything it can not arbitrarily control) and learn the right actions to do. In this section we describe a novel environment, namely some other learning agent.The first learning agent will see the actions of another learning agent and from that input try to learn the right actions. In detail, there are two ways this may happen:
\begin{enumerate}
 \item In the first mode of interaction, the second learning agent simply sees the action of the first learning agent and then goes through the learning process like the first. The only difference between the two being that the first has the percept excited with some constant external probability while the second  agent gets the percept excited with some probability that depends on the learning procedure of the first. In other words, the two agents are dealing with two different environments
 
 \item In the second mode, both agents see the same environment but the second agent gets its cue from the action of the first. This means, for example, that they both get the first percept \(p_1\) excited but the second agent rewards an action it takes only if it matches the action of the first. So for a period of time, the second agent can reward the wrong action. 

\end{enumerate}

 For this work, we assumed the second mode of interaction. An interesting question to find out is, what is the asymptotic behavior of the two learning agents for both classical and quantum learning agents.
 
 \subsection*{Discussion}
 For the quantum case, weights on the edges between vertices \((p_0, a_0) \) and \((p_1, a_1) \) do not increase uniformly together. This means that when \(p_0 \) is excited and we plot the probability of excitation of \(a_0 \) this number may be very different from the probability of excitation of \(a_1\) that is plotted when \(p_1 \) is excited.For example for the \(n^{th}\) quantum walk the probability of exciting \(a_0\) when \(p_0\) is excited might be 0.7  but on the \( (n+1)^{th}\) quantum walk if \(p_1\) is excited the probability of exciting \(a_1 \) might still be 0.25. This sort of disparities will always happen for the first few quantum walks.As a consequence, there will always be huge statistical fluctuations early in the learning process. This statistical noise reduces and steadies as the probability of doing the right action approaches the asymptotic value. It must be emphasized this noise is somewhat artificial because it comes from plotting on one graph the probability of exciting \(a_0 \) and \(a_1\) during the learning process. \newline
 
 \indent The statistical noise is made worse by higher decay rate since the learning procedure is impeded by the loss of quantum coherence. Higher decay rates also have the effect of making the agent reach its asymptotic value of learning efficiency later than the lower decay rate. \newline
 
 \indent It also turns out that the dissipative factor for the closed quantum case (first model),has less of an effect on the quantum agent than for the classical agent. This can be seen by comparing the learning efficiency for the first classical agent in the interacting case and the first quantum agent also in the interacting case. What this means is that if one can get a quantum system in which decay rates are small compared to the evolution time for the learning process, the quantum agent is more resilient to `forgetting'. This is not true for the second model of the quantum case. The dissipative factor in the learning process as a profound impact in the statistical fluctuations. Again one must emphasize that these fluctuations come from the fact the agent does not learn at the same time what right action to do is given a certain percept excitation. The agent typically learns the meaning of one percept before the other.\newline
 
  \indent What is interesting about the interacting case is that although the quantum agent looks like quantum model embedded in the classical model, the quantum agent behaves differently in the interacting environment. The second quantum agent eventually learns just as well as the original agent while the classical agent does not.This suggests that the quantum agent and the classical agent can behave differently in different environments.

\onecolumngrid
\begin{center}
	 \begin{figure}[H]
	
	   \begin{subfigure}{0.8 \textwidth}
	   	   \includegraphics[ width=\textwidth]{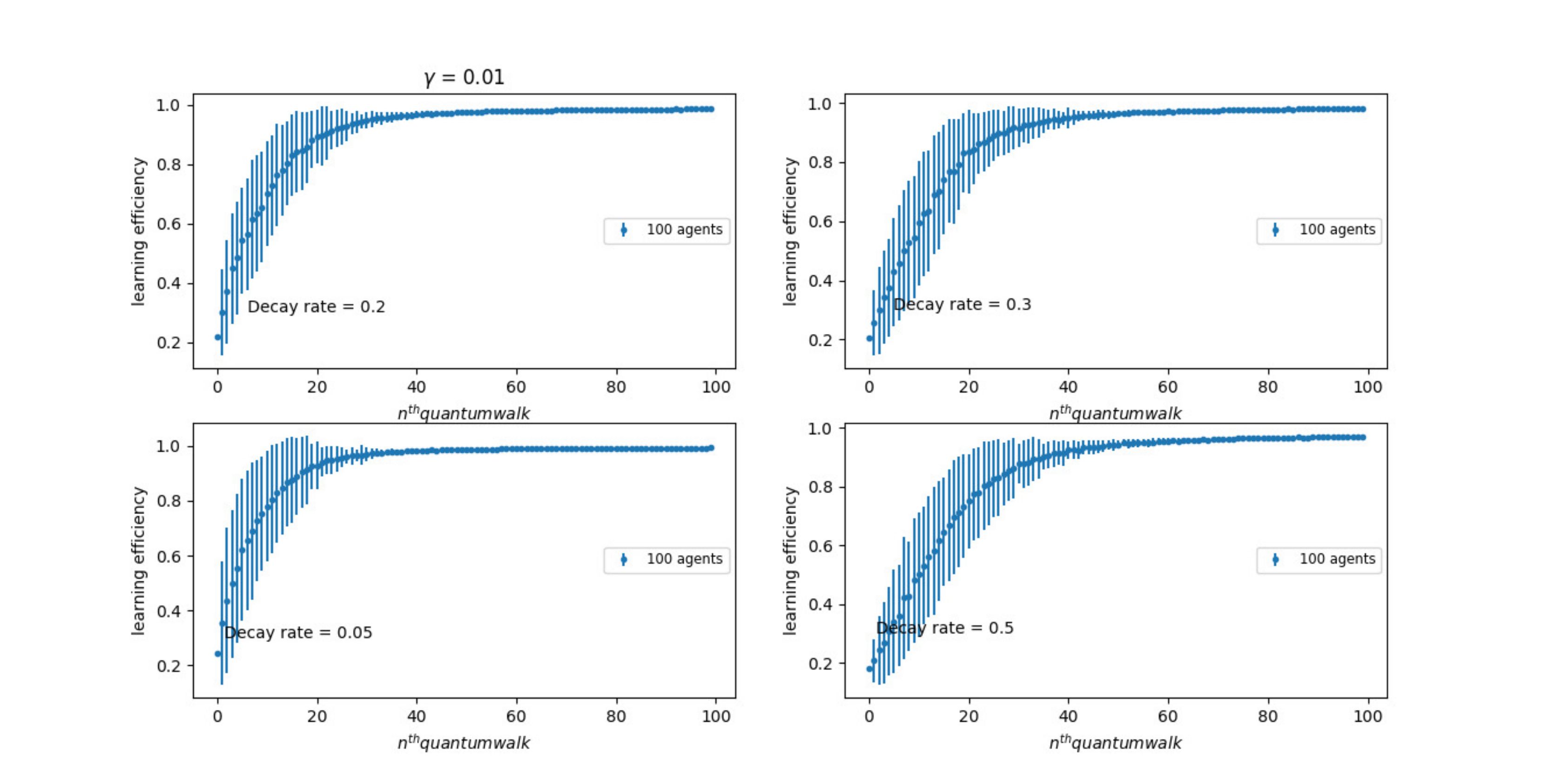}
	   \end{subfigure}
	    
		\begin{subfigure}{0.8\textwidth}
		  	\includegraphics[height=70mm, width=\textwidth]{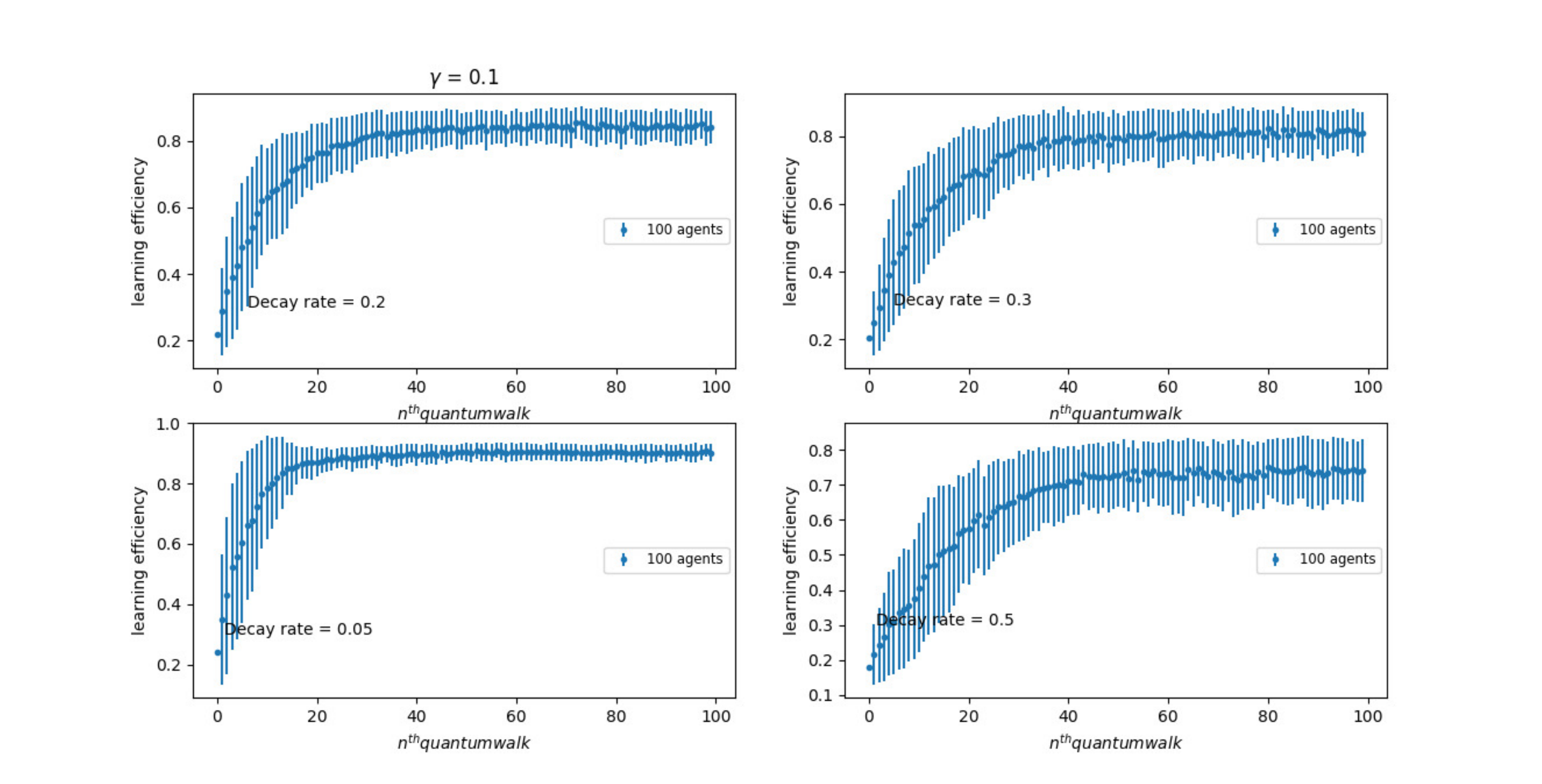}
		\end{subfigure}
	
	  \caption{First quantum model}
	 \end{figure}
\end{center}

\begin{center}

 \begin{figure}
   \centering
   \begin{subfigure}{0.9\textwidth}
     \includegraphics[width=\textwidth]{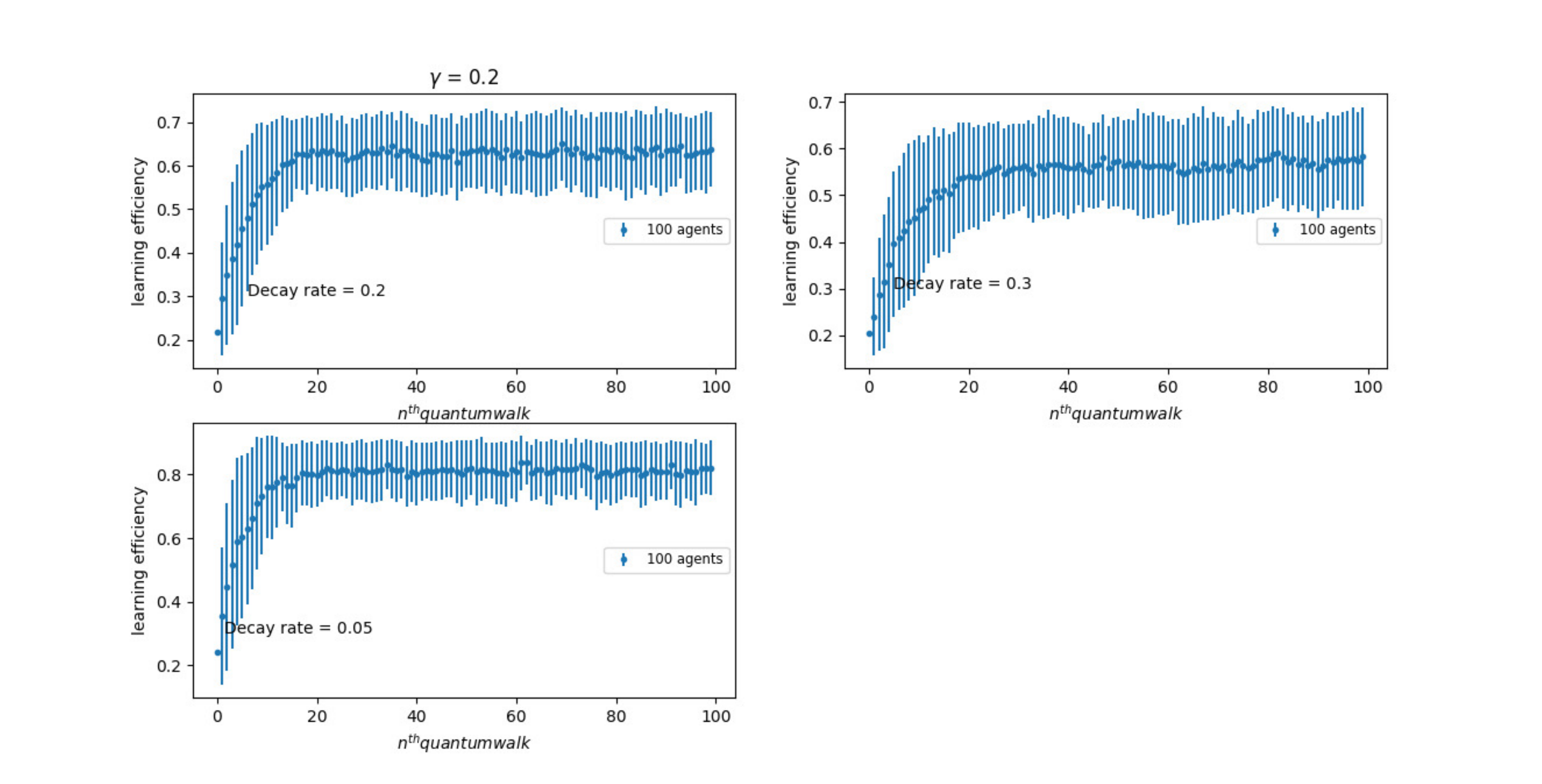}
     \caption{First quantum model}
   \end{subfigure}
		 	 
	 \begin{subfigure}{0.9\textwidth}
	 	 \centering
	  	\includegraphics[width=\textwidth]{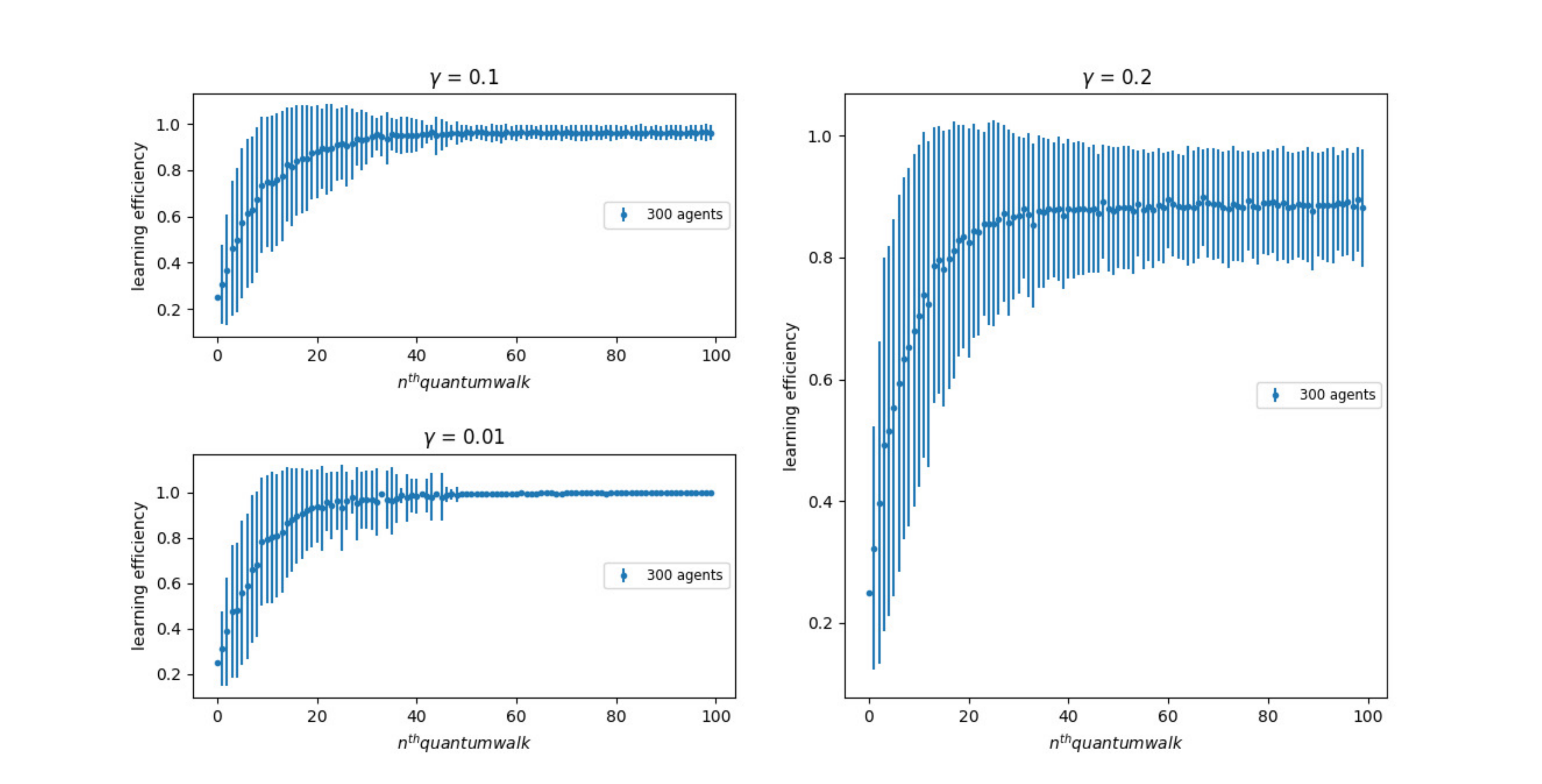}
	  	\caption{Second quantum model(no noise)}
	 \end{subfigure}
\end{figure}
\end{center}

\twocolumngrid

 \onecolumngrid
 \begin{figure}[H]
  \centering
  \begin{subfigure}[b]{0.9\textwidth}
	 \includegraphics[ width=\textwidth]{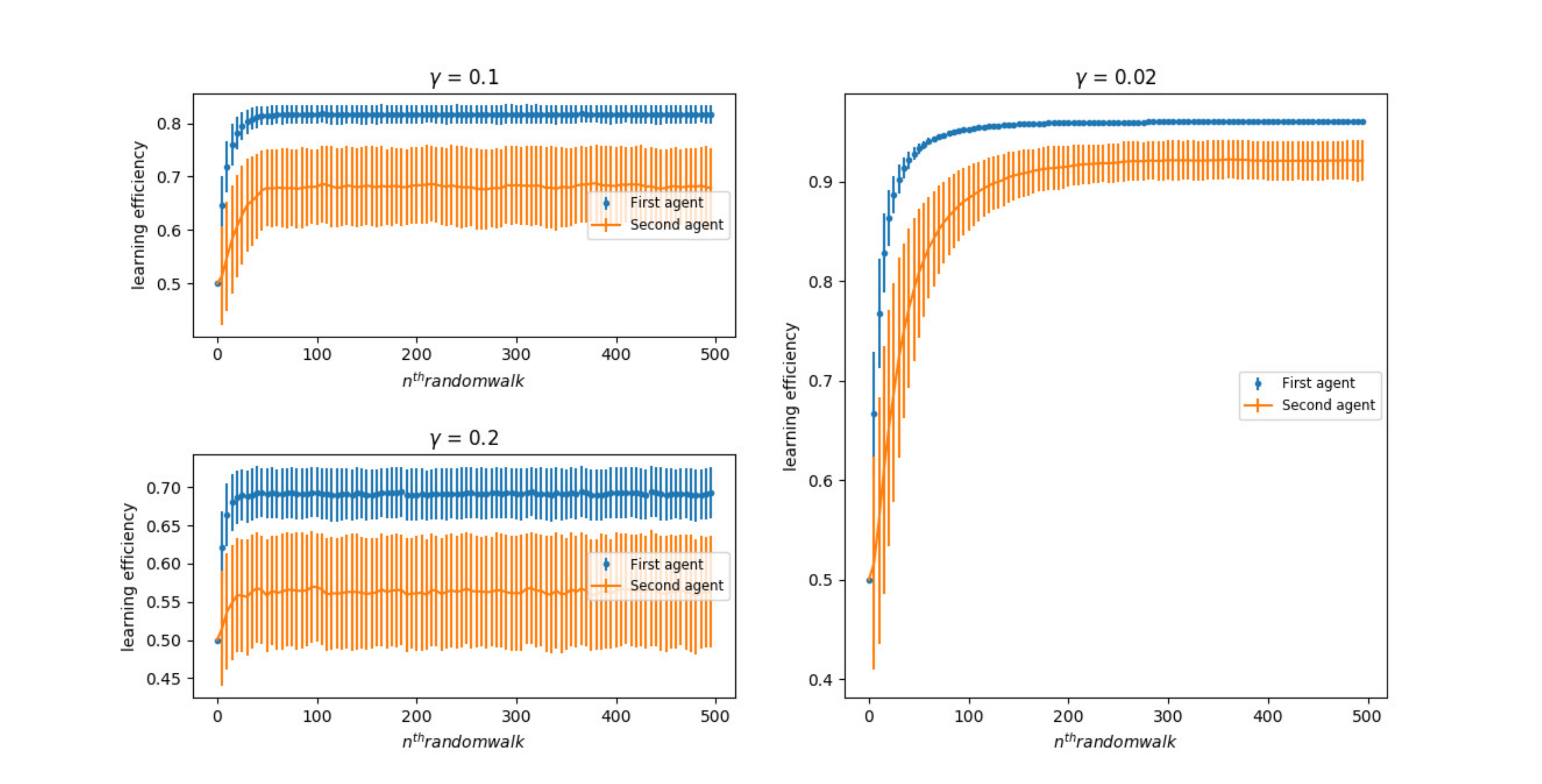}
	\caption{Classical interacting agents, Averaged over 1000 agents} 
  \end{subfigure}
  \centering
  \begin{subfigure}[b]{0.9 \textwidth}
 \includegraphics[width= \textwidth]{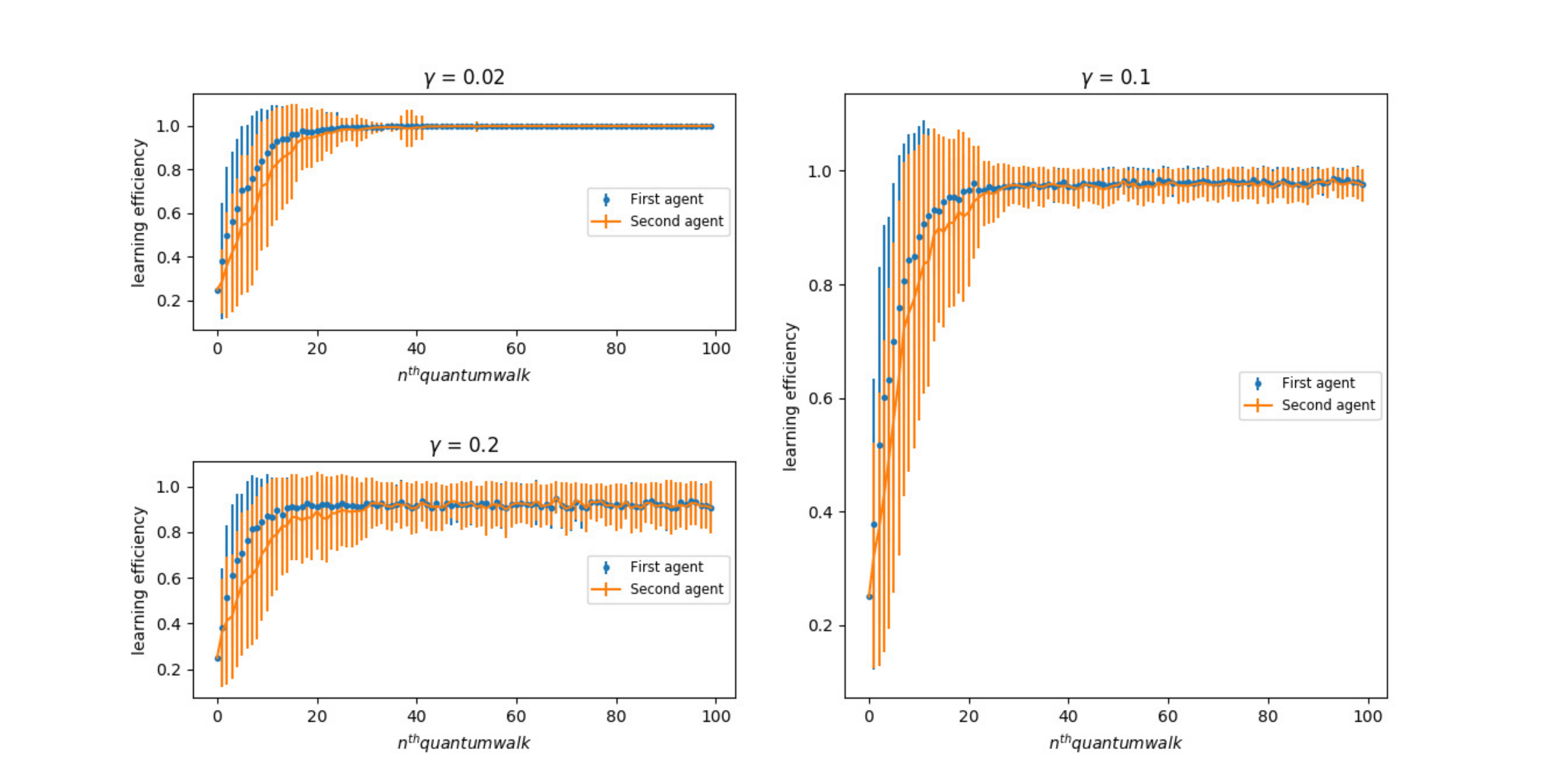}
 \caption{Quantum interacting agents (no noise), Averaged over 100 agents}
  \end{subfigure}
   
 \end{figure}
 
 \twocolumngrid
  \section{Conclusion}
  We have presented two models of the quantum projective agent and shown their ability to learn. The quantum model's performance is not only affected by the dissipative factor in the learning process but also by the decay of the quantum state. Decay rates of the quantum state have a secondary effect on the learning efficiency but give a larger spread in the statistical noise. The statistical noise comes mostly from the fact that the quantum agent does not learn what all the symbols mean at the same time. We have also seen that the learning procedure can be impacted by the Hamiltonian used since the second model was more sensitive to the dissipative factor. We introduced a setting in which the classical learning agent is not equivalent to the quantum agent. This shows it is worth studying the quantum case for other reasons apart from getting speed ups. 

\bibliographystyle{plain}
\bibliography{MasterBibliography}

\begin{thebibliography}{10}

\bibitem{Brown2016}
Kenneth~R. Brown, Jungsang Kim, and Christopher Monroe.
\newblock Co-designing a scalable quantum computer with trapped atomic ions.
\newblock {\em Quantum Information}, 2(16034), November 2016.

\bibitem{Friis2015}
Nicolai Friis, Alexey~A. Melnikov, Gerhard Kirchmair, and Hans J.Briegel.
\newblock Coherent controlization using superconducting qubits.
\newblock {\em Scientific Reports}, 5(18036), Decemeber 2015.

\bibitem{J.Briegel2012}
Hans J.Briegel and Gemma~De las Cuevas.
\newblock Projective simulation for artificial intelligence.
\newblock {\em Scientific Reports}, 2(400), May 2012.

\bibitem{Johansson2013}
J.R. Johansson, P.D. Nation, and F.Nori.
\newblock Qutip 2: A python framework for the dynamics of open quantum systems.
\newblock {\em Comp. Phys. Comm}, 184(1234), 2013.

\bibitem{Melnikov2015}
Alexey~A. Melnikov, Adi Makmal, Vedran Dunjko, and Hans J.Briegel.
\newblock Projective simulation with generalizaton.
\newblock {\em preprint arXiv:1504.02247v1}, 2015.

\bibitem{M.Tiersch2015}
M.Tiersch, E.J.Ganahl, and H.J.Briegel.
\newblock Adaptive quantum computation in changing environments using
  projective simulation.
\newblock {\em Scientific Reports}, 5(12874), August 2015.

\bibitem{Paparo2014}
Guiseppe~Davide Paparo, Verfran Dunjko, Adi Makmal, Miguel~Angel
  Martin-Delgado, and Hans J.Briegel.
\newblock Quantum speedup for active learning agents.
\newblock {\em Physical Review X}, 4:14, July 2014.

\bibitem{Sutton1998}
Richard~S. Sutton and Andrew~G. Barto.
\newblock {\em Reinforcement Learning: An Introduction (Adaptive Computation
  and Machine Learning)}.
\newblock A Bradford Book, 1 edition, March 1998.

\bibitem{Szegedy2004}
M.~Szegedy.
\newblock Quantum speed-up of markov chain based algorithms.
\newblock {\em Foundation of Computer Science, 2004. Proceedings. 45th Annual
  IEEE Symposium on Foundations of Computer Science}, December 2004.

\bibitem{Viamontes2005}
George~F. Viamontes, Igor~L. Markov, and John~P. Hayes.
\newblock Is quantum search practical?
\newblock {\em Computing in Science \& Engineering}, 7(Issue 3), May-June 2005.

\end{thebibliography}
\end{document}